\documentclass[aps,prd,superscriptaddress,twocolumn,nofootinbib]{revtex4-2}
\usepackage[english]{babel}
\usepackage[utf8]{inputenc}
\usepackage[colorinlistoftodos, color=green!40, prependcaption]{todonotes}
\usepackage[pdftex, pdftitle={Article}, pdfauthor={Author}]{hyperref} % For hyperlinks in the PDF

\hypersetup{colorlinks=true, allcolors=blue}
\usepackage[english]{babel}
\usepackage{xcolor}
\usepackage{graphicx}
\usepackage{amssymb}
\usepackage{amsmath}
\usepackage{slashed}
\usepackage{xcolor}
\usepackage{mathtools}
\usepackage{hyperref}
\usepackage{floatrow}
\usepackage{makecell}
\usepackage{multirow}
\usepackage{booktabs}
\usepackage{enumitem}
\usepackage{placeins}
\usepackage{arydshln}
\usepackage{cancel}
\usepackage{natbib}
\usepackage{soul}
\usepackage{orcidlink}

\providecommand{\sorthelp}[1]{}

\newcommand{\lm}{{\ell m}}
\newcommand{\ch}{\mathrm{c}}
\newcommand{\chp}{\mathrm{c}_1}
\newcommand{\chs}{\mathrm{c}_2}
\newcommand{\ncmb}{\mathrm{CMB}}
\newcommand{\nfg}{\mathrm{FG}}
\newcommand{\ntot}{\mathrm{SKY}}
\newcommand{\nfac}{\mathcal{F}}
\newcommand{\ncal}{\mathcal{K}}
\newcommand{\blnu}{b_{\ell,\nu}}
\newcommand{\bandchnu}{\tau^\ch_\nu}
\newcommand{\Planck}{\textit{Planck}}

\begin{document}
\title{Modeling beam chromaticity for high-resolution CMB analyses}

\author{S. Giardiello~\orcidlink{0000-0002-8340-3715}}~\email{GiardielloS@cardiff.ac.uk}~\affiliation{School of Physics and Astronomy, Cardiff University, The Parade, CF24 3AA Cardiff, Wales, UK}

\author{A. J. Duivenvoorden~\orcidlink{0000-0003-2856-2382}}~\affiliation{Max-Planck-Institut f{\"u}r Astrophysik, Karl-Schwarzschild Str. 1, 85741 Garching, Germany}~\affiliation{Center for Computational Astrophysics, Flatiron Institute, 162 5th Avenue, New York, NY 10010 USA}~\affiliation{Joseph Henry Laboratories of Physics, Jadwin Hall,
Princeton University, Princeton, NJ, USA 08544}

\author{E. Calabrese~\orcidlink{0000-0002-8340-3715}}~\affiliation{School of Physics and Astronomy, Cardiff University, The Parade, CF24 3AA Cardiff, Wales, UK}

\author{G. Galloni~\orcidlink{0000-0003-0837-0068}}~\affiliation{Dipartimento di Fisica e Scienze della Terra, Universit\`a degli Studi di Ferrara, via Saragat 1, I-44122 Ferrara, Italy}~\affiliation{Istituto Nazionale di Fisica Nucleare, Sezione di Ferrara, via Saragat 1, I-44122 Ferrara, Italy}

\author{M. Hasselfield~\orcidlink{0000-0002-2408-9201}}~\affiliation{Center for Computation Astrophysics, Flatiron Institute, 162 5th Ave 9th floor, New York, NY 10010}

\author{J.~C. Hill~\orcidlink{0000-0002-9539-0835}}~\affiliation{Department of Physics, Columbia University, New York, NY 10027, USA}

\author{A. La Posta~\orcidlink{0000-0002-2613-2445}}~\affiliation{Department of Physics, University of Oxford, Denys Wilkinson Building, Keble Road, Oxford OX1 3RH, United Kingdom}

\author{T. Louis~\orcidlink{0000-0002-6849-4217}}~\affiliation{Université Paris-Saclay, CNRS/IN2P3, IJCLab, 91405 Orsay, France"}

\author{M. Madhavacheril~\orcidlink{0000-0001-6740-5350}}~\affiliation{Department of Physics and Astronomy, University of Pennsylvania,
209 South 33rd Street, Philadelphia, PA, USA 19104}

\author{L. Pagano~\orcidlink{0000-0003-1820-5998}}~\affiliation{Dipartimento di Fisica e Scienze della Terra, Universit\`a degli Studi di Ferrara, via Saragat 1, I-44122 Ferrara, Italy}~\affiliation{Istituto Nazionale di Fisica Nucleare, Sezione di Ferrara, via Saragat 1, I-44122 Ferrara, Italy}~\affiliation{Institut d'Astrophysique Spatiale, CNRS, Univ. Paris-Sud, Universit\'{e} Paris-Saclay, B\^{a}t. 121, 91405 Orsay cedex, France}

\date{\today} % Leave empty to omit a date

\begin{abstract}
We investigate the impact of beam chromaticity, i.e., the frequency dependence of the beam window function, on cosmological and astrophysical parameter constraints from CMB power spectrum observations. We show that for future high-resolution CMB measurements it is necessary to include a color-corrected beam for each sky component with a distinct spectral energy distribution. We introduce a formalism able to easily implement the beam chromaticity in CMB power spectrum likelihood analyses and run a case study using a Simons Observatory (SO) Large Aperture Telescope-like experimental setup and within the public SO software stack. To quantify the impact, we assume that beam chromaticity is present in simulated spectra but omitted in the likelihood analysis. We find that, for passbands of fractional width $\Delta \nu/\nu \sim 0.2$, neglecting this effect leads to significant biases, with astrophysical foreground parameters shifting by more than $2\sigma$ and cosmological parameters by significant fractions of the error.
\end{abstract}


\maketitle

\section{Introduction} \label{sec:intro}

The increase in precision in Cosmic Microwave Background (CMB) observations which we have witnessed over the last three decades~\citep[see e.g.,][]{WMAP:2012fli,WMAP:2012nax,Planck:2018nkj,planck2018_params,ACT:2020gnv,ACT:2020frw,SPT-3G:2021eoc,SPT-3G:2022hvq,POLARBEAR:2014hgp, BICEP:2021xfz} has demanded a large effort in developing data analysis pipelines able to account for precise characterization of the instruments and of the sky emission. This work is essential to make sure that the very tight constraints set on cosmological models are robust and unbiased. For example, the subsequent releases of the \href{https://www.esa.int/Science_Exploration/Space_Science/Planck}{\Planck} mission data (from the initial early survey results to the latest NPIPE products) have seen the deployment of many techniques to reduce systematics arising from e.g., uncertain instrument performance or the scanning strategy, and contamination from Galactic emission~\cite{Zacchei:2011ba,PlanckHFICoreTeam:2011bh,Planck:2015qep,Planck:2013cta,Planck:2015qep,Planck:2015aiq,Planck:2015hzl,Planck:2018bsf,Planck:2018lkk,planck_npipe,Planck:2013win,Planck:2015bpv,Planck:2019nip,Efstathiou:2019mdh,Delouis:2019bub}. While the \Planck\ satellite focused on getting the most robust large scale modes ($\ell \lesssim 1500-2000$), from the ground, experiments like the Atacama Cosmology Telescope \href{https://act.princeton.edu/}{(ACT)} and the South Pole Telescope \href{https://pole.uchicago.edu/public/Home.html}{(SPT)} have refined intermediate and small angular scales ($1000\lesssim \ell \lesssim 10000$). The requirement for those scales is to tackle in particular the astrophysical foreground emission from extra-galactic sources and other unresolved signals~\cite{Dunkley:2013vu,ACT:2020frw,George:2014oba,SPT:2020psp}, and the frequency-dependent systematics effects which couple with them. If ignored, these systematics can lead to incorrect estimates of foreground parameters -- preventing their astrophysical and cosmological interpretation~\cite{Komatsu:2002wc, Madhavacheril:2017onh,Douspis:2021ing,Planck:2013wqd,Planck:2015emq,Maniyar:2020tzw,Zagatti:2024bhs}-- and potentially introduce biases in the estimation of cosmological parameters. Next-generation ground-based experiments like the Simons Observatory (SO)~\cite{SimonsObservatory:2018koc} and CMB-S4~\cite{Abazajian:2013bxd} are actively developing these aspects of the analysis pipeline~\cite{Abitbol:2020fvn, ACT:2023wcq, Planck:2015zbi,Giardiello:2024uzz,Dachlythra:inprep}.

A crucial ingredient for a correct estimate of CMB and foreground emission is the accurate knowledge of the beam, which represents the optical response of the instrument. Its width encodes the resolution of the instrument and its azimuthal and polar profile depends on the entire optical chain of the instrument. It is customary, in particular for high-resolution analyses, to use the assumption of azimuthally-symmetric beams~\cite{Challinor:2000xy}, which holds well for the main beam component and simplifies the beam to its radial profile $b(\theta)$, function of the polar angle $\theta$. This works in the case of a redundant scanning strategy, where each pixel is observed from many angles and the averaged beam gets symmetrized.

Usually beam profiles are estimated employing observations of planets or known sources \cite{Huffenberger:2010ew,Lungu:2021slc} and, for broad passbands, the resulting beams can strongly depend on source spectral type. Thus, the effective beam profile may vary for different sources of emissions. However, in many previous cosmological analyses this effect has often been neglected~\cite{Planck:2019nip, SPT:2017jdf} or not included in the baseline analysis because too small to significantly impact cosmological parameter inference\footnote{The ACT DR4 analysis computed an approximate color-correction to adjust the beam estimated from Uranus observations to the beam appropriate for the CMB and used those in the baseline likelihood. Color-corrections for the other sky components were neglected. Tests assessing the impact of beam chromaticity were done as a robustness test of the cosmology results~\cite{ACT:2020frw}; finding negligible impact on cosmological parameters and O$(\lesssim1\sigma)$ in astrophysical parameters, the extended beam modeling was not used in the baseline likelihood.}~\cite{ACT:2020gnv, ACT:2020frw}, particularly given the sensitivity or the multipole range of the experiment.
Only some recent component separation studies~\cite{Madhavacheril:2019nfz, ACT:2023wcq} included this effect for the first time in the baseline analysis settings. As we show in this paper, with the increasing sensitivity of upcoming experiments, it will be essential to accurately measure and incorporate the frequency dependence of beam profiles into the analysis pipeline. 
This will allow proper modeling of the observed power spectra and ensure an unbiased recovery of both cosmological and foreground parameters.

In this work, we lay out the formalism for the integration of beam chromaticity in the power spectrum and likelihood analysis of a CMB experiment, and show the potential impact of this effect on the recovery of parameters from an SO Large Aperture Telescope (LAT)-like experiment. We derive the mathematics needed for the integration of this term in the calculation of the foreground Spectral Energy Distributions (SEDs), and we implement the modeling of the chromatic beams in the public SO power spectrum likelihood code \texttt{LAT\_MFLike}\footnote{\href{https://github.com/simonsobs/LAT_MFLike/releases/tag/v1.0.0}{\texttt{LAT\_MFLike}, version 1.0.0}} and its foreground spectrum library \texttt{fgspectra}\footnote{\href{https://github.com/simonsobs/fgspectra/releases/tag/v1.3.0}{\texttt{fgspectra}, version 1.3.0}}. 

The paper is organized as follows. We describe our formalism in Section~\ref{sec:method}, then present its implementation and results on how the chromatic beam effect can bias cosmological and foreground parameters in Section~\ref{sec:implementation}. We then draw conclusions in Section~\ref{sec:conclusions}.

\section{Methodology}\label{sec:method}
For an ideal monochromatic observation, the beam window function can be simply computed as the Legendre polynomial transform of the frequency-dependent radial profile $b(\theta,\nu)$~\cite{Lungu:2021slc}: 
\begin{equation}
\blnu \propto \int^{+1}_{-1}d\cos{\theta} P_\ell(\cos{\theta})b(\theta,\nu),
\end{equation}
where $P_\ell(\cos{\theta})$ are the Legendre polynomials. More realistically, if we consider a broad passband, a map of the microwave sky (measuring CMB and foreground emission) in harmonic space, for a frequency channel $\ch$, reads: 

\begin{align} \label{eq:a_tot}
a_\lm^{\ntot,\ch}&=a_\lm^{\ncmb,\ch}+a_\lm^{\nfg,\ch}=\nonumber\\
&=\frac{1}{\ncal^\ch}\int d\nu \, \bandchnu\,\nfac_\nu\,\blnu^c \left(a_\lm^\ncmb + a_{\lm,\nu}^{\nfg}\right)=\nonumber\\
&=b_{\ell}^{\ncmb,\ch}a_\lm^\ncmb + \frac{1}{\ncal^\ch}\int d\nu\, \bandchnu\,\nfac_\nu\, \blnu^\ch \, a_{\lm,\nu}^\nfg,
\end{align}

\noindent where $\bandchnu$ is the passband\footnote{Throughout this work we assume that the passband $\bandchnu$ is proportional to the instrument's response to a beam-filling source with a frequency-independent surface brightness ($\propto \nu^0$).} of the channel, $\nfac_\nu$ is the differential temperature to surface brightness unit conversion factor, defined as:
\begin{align*}
   \nfac_\nu &\equiv\left.\frac{\partial B(T,\nu)}{\partial T}\right|_{T=T_{\ncmb}} = \\
   & =\left( \frac{2 k_B}{c^2} \right) \nu^2 \frac{x^2 e^x}{(e^x-1)^2} \, \left[ \frac{\mathrm{W}}{\mathrm{m^2 \, sr \, Hz \, K}} \right],    
\end{align*}
\noindent where $B(T,\nu)$ is the Planck function and $x\equiv h\nu/k_{B}T_{CMB}$.
$\ncal^\ch$ is defined as:

\begin{equation}
   \ncal^\ch = \int d\nu\, \bandchnu\, \nfac_\nu,
\end{equation}

\noindent and effectively acts as a calibration factor bringing the map from surface brightness back to differential temperature units. This is a consequence of our choice of calibrating the map on the CMB. Finally, the beam window function for channel c, $\blnu^\ch$, is normalized to 1 at $\ell = 0$.
In Eq.~\ref{eq:a_tot} we are assuming that both CMB and foregrounds are in CMB differential temperature units, so they have the same conversion factors to surface brightness units and $a_\lm^{\ncmb,\ch}$ has no frequency dependence\footnote{The code used to simulate the foreground power spectra, \href{https://github.com/simonsobs/fgspectra/tree/main}{\texttt{fgspectra}}, simulates the foreground SEDs in Rayleigh-Jeans (RJ) units and then multiplies them by a conversion factor bringing RJ to CMB units:
\begin{equation}
    t_\nu \propto \frac{(e^x - 1) ^ 2}{x^2 e^x}, \quad x = \frac{h \nu}{k_B T_{CMB}},
\end{equation}
such that the final SED is simulated in CMB differential temperature units. The difference between the SEDs in RJ units and the ones simulated in surface brightness units is a $\nu^2$ factor: $f^\mathrm{FG}_\mathrm{SB} \propto f^\mathrm{FG}_\mathrm{RJ} \nu^2$.}.

The window function $b_{\ell}^{\ncmb,\ch}$ appearing in Eq.~\ref{eq:a_tot} is referred to as  ``CMB" beam  (i.e. the beam obtained assuming flat emission in differential temperature units):
\begin{equation}
   b_{\ell}^{\ncmb,\ch} = \frac{1}{\ncal^\ch}\int d\nu \, \bandchnu \, \nfac_\nu \, \blnu^\ch.
\end{equation}

Indeed, in the presence of CMB-only emission, the input sky $a_\lm$ factors out from the integral as in the first term of Eq.~\ref{eq:a_tot}. However, this factorization does not hold when foreground emission is present. To model the signal from foregrounds, we follow Ref.~\cite{Tegmark:1999ke, Dunkley:2013vu} and separate it into a spectral energy distribution (SED) and an angular-dependent component, defined at a pivot frequency $\nu_0$:

\begin{equation} \label{eq:aFG}
a_{\ell m}^{\nfg}(\nu) = f^\mathrm{FG}_{\nu,\nu_0} a^\mathrm{FG}_{\ell m}(\nu_0)
\end{equation}
(for brevity, we drop $\nu_0$ in the following). Hence the cross spectrum of channels $\chp$ and $\chs$ reads:

\begin{align} \label{eq:C_tot}
&C_{\ell}^{\ntot, \mathrm{c}_1 \mathrm{c}_2}
= C_{\ell}^{\ncmb} b_{\ell}^{\ncmb, \chp} b_{\ell}^{\ncmb,\chs} + \nonumber \\
&+ \frac{\left[\int d\nu \, \tau^{\mathrm{c}_1}_{\nu} \, \nfac_{\nu} b^{\chp}_{\ell, \nu}\, f^\mathrm{FG}_{\nu}\right]}{\ncal^{\mathrm{c}_1}} \frac{\left[\int d\nu \, \tau^{\mathrm{c}_2}_{\nu} \, \nfac_{\nu} b^{\chs}_{\ell, \nu} \, f^\mathrm{FG}_{\nu}\right]}{ \ncal^{\mathrm{c}_2}}C^{\nfg}_{\ell}.
\end{align}

This expression represents the power spectrum still convolved by the beam window function. We can thus factorize the CMB beams and deconvolve them, obtaining:
\begin{align} \label{eq:C_tot_bfact}
&\hat{C}_{\ell}^{\ntot, \mathrm{c}_1 \mathrm{c}_2}= C_{\ell}^{\ncmb} + \nonumber\\
&+ \frac{\left[\int d\nu \, \tau^{\chp}_{\nu} \, \nfac_{\nu} b^{\chp}_{\ell, \nu}\,f^\mathrm{FG}_{\nu} \right]}{\left[  \int d\nu \, \tau^{\chp}_{\nu} \, \nfac_{\nu} b^{\chp}_{\ell, \nu}\right]} \frac{\left[ \int d\nu \, \tau^{\chs}_{\nu} \, \nfac_{\nu} b^{\chs}_{\ell, \nu} \,f^\mathrm{FG}_{\nu}\right]}{\left[ \int d\nu \, \tau^{\chs}_{\nu} \, \nfac_{\nu} b^{\chs}_{\ell, \nu} \right]}C^{\nfg}_{\ell}.
\end{align}

Now this expression can be generalized to any number of different foreground components, with the total foreground power spectrum, already deconvolved by the CMB beams, given by:

\begin{align} \label{eq:clFG}
\hat{C}_{\ell}^{\nfg_{ij},\mathrm{c}_1 \mathrm{c}_2}&=\sum_{\nfg_{ij}}  C^{\nfg_{ij}}_{\ell} \int d\nu \, r_{\ell,\nu}^{\mathrm{c}_1} f^{\mathrm{FG}_i}_{\nu} \int d\nu\,  r_{\ell,\nu}^{\mathrm{c}_2} \, f^{\mathrm{FG}_j}_{\nu} = \nonumber \\ 
&=\sum_{\nfg_{ij}}  C^{\nfg_{ij}}_{\ell} \,\hat{f}_\ell^\mathrm{FG_i, \chp} \hat{f}_\ell^\mathrm{FG_j, \chs}
\end{align}

\noindent where we have defined the channel-dependent geometric factor:

\begin{equation} \label{eq:rcomplete}
    r_{\ell,\nu}^{\ch} = \frac{\bandchnu \, \nfac_{\nu} \, \blnu^\ch}{\int d\nu \, \bandchnu \, \nfac_{\nu} \, \blnu^\ch} .
\end{equation}
\noindent and the integral:
\begin{equation} \label{eq:fhat}
    \hat{f}_\ell^\mathrm{FG, \ch} = \int d\nu \, r_{\ell,\nu}^{\ch} f^\mathrm{FG}_{\nu}.
\end{equation}

The indices of the foreground components $\mathrm{FG}_i, \mathrm{FG}_j$ mean that the power spectra can be computed also between different components.

Equation~\ref{eq:clFG} represents the power spectrum of the total foreground emission distorted by the chromaticity of the main beam.

It is now instructive to show what happens if we assume achromatic beams. In this case the beam window function factorizes out and the sky $a_{\ell m}$ reads:

\begin{equation}
\tilde{a}_{\lm}^{\ntot,\ch}=b_{\ell}^{\ncmb,\ch}\left(a_{\lm}^{\ncmb} + \frac{1}{\ncal^{\ch}} \int d\nu \, \bandchnu \, \nfac_\nu  \, a_{\lm, \nu}^{\nfg}\right),
\end{equation}
where $\tilde{}$~is used to represent the case with achromatic beams.

The foreground power spectrum (already deconvolved by the CMB beam) is then given by:

\begin{align}\label{eq:clFGnochormaticity}
&\hat{\tilde{C}}_{\ell}^{\nfg,\mathrm{c}_1 \mathrm{c}_2}=
&\sum_\nfg C^{\nfg}_{\ell} \int d\nu \, \tilde{r}_{\nu}^{\chp} f^\mathrm{FG}_{\nu} \int d\nu \, \tilde{r}_{\nu}^{\chs} \, f^\mathrm{FG}_{\nu},
\end{align}

\noindent with
\begin{equation} \label{eq:rtilde}
    \tilde{r}_{\nu}^{\ch} = \frac{\bandchnu \, \nfac_{\nu}}{\ncal^\ch},
\end{equation}

\noindent which reduces to simple bandpass integration of the foreground emission~\cite{Abitbol:2020fvn}.

The formalism outlined so far can easily be generalized to both temperature and polarization, simply by considering their corresponding beams, i.e., $b^{\mathrm{T/E/B}, \ch}_{\ell}$. 
In general, distinguishing between T and E geometrical factors, we have:

\begin{align}
\hat{C}_{\ell}^{\nfg,\textrm{TE},\mathrm{c}_1 \mathrm{c}_2}&= C^{\nfg, \mathrm{TE}}_{\ell} \int d\nu \, r_{\ell, \nu}^{\mathrm{T}, \chp} f^\mathrm{FG, \mathrm{T}}_{\nu} \int d\nu \, r_{\ell, \nu}^{\mathrm{E}, \chs} \, f^\mathrm{FG, \mathrm{E}}_{\nu}.
\end{align}

Aside from the assumption of azimuthal symmetry, we have maintained complete generality in the definition of the beam until now.

The main goal of our work is to understand if ignoring beam chromaticity can have an impact -- and if yes, how much -- on cosmological and foreground parameter estimates from CMB power spectrum data. In the formalism that we have just outlined, this can be done by simply comparing the foreground power spectra computed with Eq.~\ref{eq:clFG}, i.e., with beam chromaticity, and Eq.~\ref{eq:clFGnochormaticity}, i.e., neglecting its effect. The essential elements of this comparison are the chromatic beam window functions, $b_{\ell,\nu}$. Estimating these beams is a complex task, typically requiring a combination of planet observations and detailed optical simulations \cite{Duivenvoorden:inprep, Huffenberger:2010ew}.

In our analysis, for simplicity, we can assume diffraction-limited Gaussian beams for each spectral element. Under this assumption, the Full Width Half Maximum (\texttt{FWHM}) for a frequency $\nu$ is simply given by

\begin{equation}
   \texttt{FWHM}(\nu) \sim \frac{1}{D}\frac{c}{\nu}\; [\texttt{rad}],
\end{equation}
where $D$ is the aperture of the primary mirror of a telescope\footnote{This is obtained by fitting a Gaussian profile to the Fraunhofer diffraction pattern produced by a circular aperture.}. For a Gaussian beam the window function can then be computed as~\cite{Challinor:2000xy}:
\begin{equation}\label{eq:gaussbl}
b_{\ell, \nu} = \exp\left(-\frac{1}{2}\ell (\ell+1) \left(\frac{\texttt{FWHM}(\nu)}{2 \sqrt{2 \ln 2}}\right)^2\right).
\end{equation}

In order to further generalize this expression we can add an extra frequency scaling \cite{Duivenvoorden:inprep} as:
\begin{equation} \label{eq:nu_scaling}
   \texttt{FWHM}(\nu)=\texttt{FWHM}(\nu_0)\left(\frac{\nu}{\nu_0}\right)^{-\alpha/2} = \frac{1}{D}\frac{c}{\nu_0} \left(\frac{\nu}{\nu_0}\right)^{-\alpha/2},
\end{equation}
where the case $\alpha = 2$ corresponds to a diffraction-limited experiment and $\alpha = 0$ neglects any chromaticity effect. 

\begin{figure}[t!]
 \includegraphics[width=0.8\textwidth]{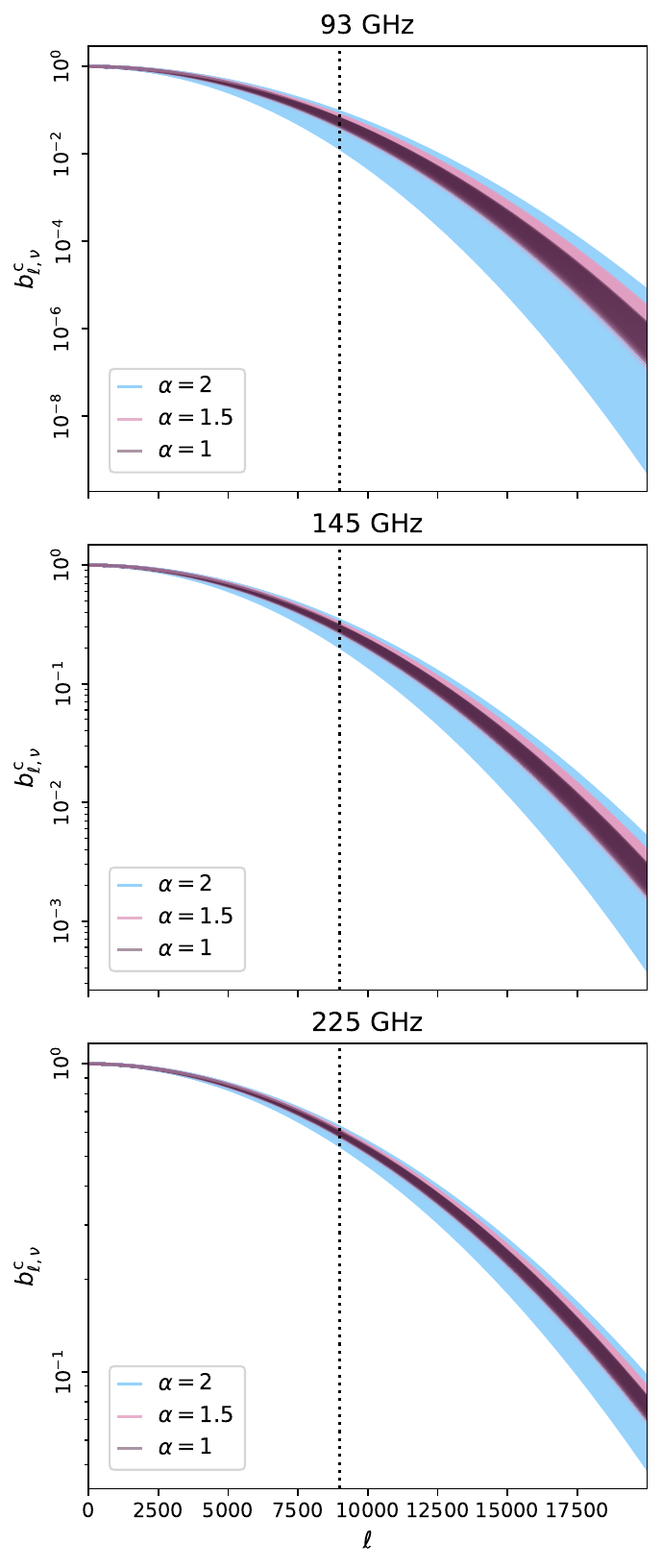}
 \caption{Variation of Gaussian beam profiles, $\blnu^\ch$, with $\alpha$ for three cosmological channels, $\ch =$ 93, 145 and 225 GHz, (for e.g., the SO LAT). The \texttt{FWHM} is computed as in Eq.~\ref{eq:nu_scaling}, using a diameter of $D$ = 6 m (characteristic of CMB experiments targeting high-resolution, arcminute-scale measurements). The $\blnu$s are plotted for $\sim 30$ GHz-wide passbands. Since the bandpass fractional width $\Delta \nu/\nu$ is the largest for channel 93 GHz, the variation in frequency of the chromatic beams is larger for this channel than the other ones. The dotted line shows the maximum multipole ($\ell = 9000$) considered in this analysis and typical of multi-frequency CMB measurements.} \label{fig:bl_nu}
\end{figure}

\begin{figure}[t!]
 	\includegraphics[width=0.8\textwidth]{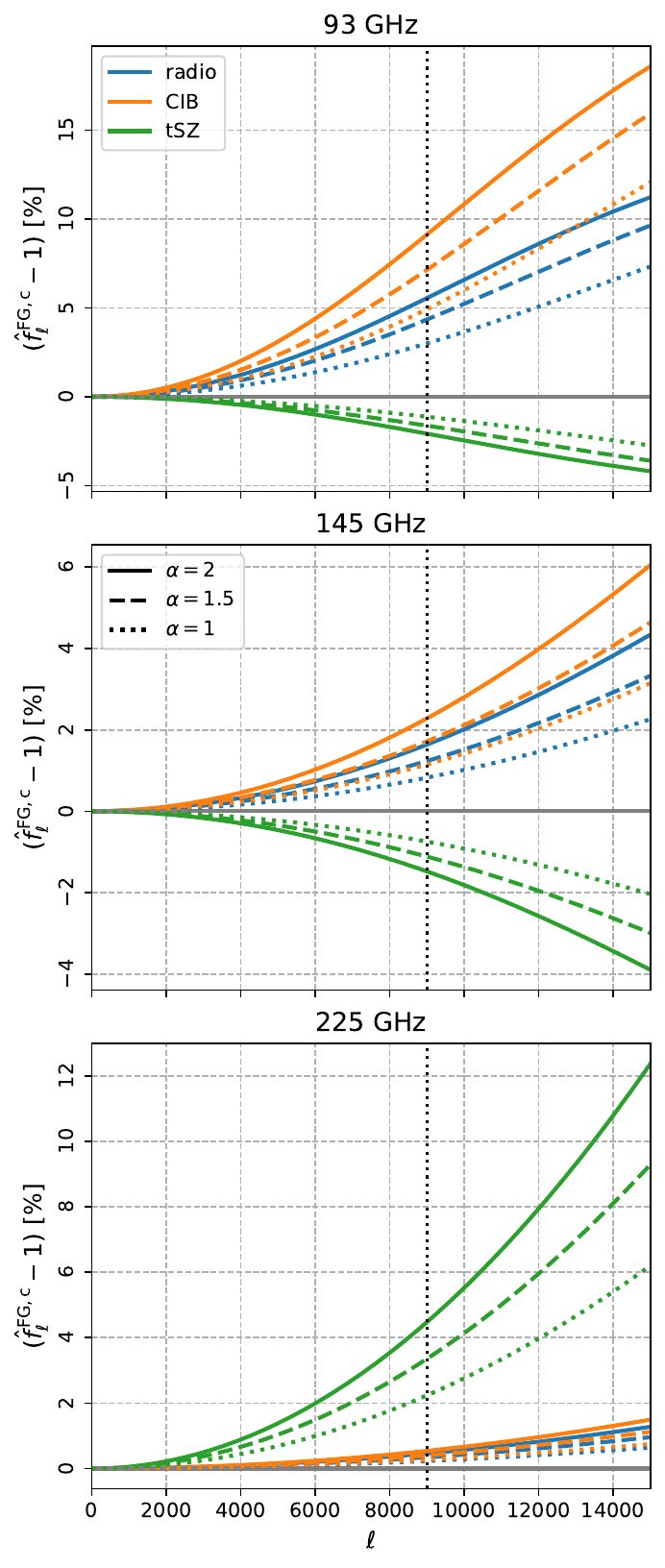}
 	\caption{Fractional variation of the foreground SED integrated in frequency with the normalized beam and passband, i.e. $(\hat{f}^\mathrm{FG, \ch}_\ell -1)$ , where $\hat{f}^\mathrm{FG,\ch}_\ell$ is defined in Eq.~\ref{eq:fhat}. Each passband has a width of $\sim 30$ GHz and we assume a top-hat passband for simplicity. We are plotting the temperature $f^\mathrm{FG}_\mathrm{SED}$ of radio (blue), CIB (orange) and tSZ (green) integrated with Gaussian beams with $\alpha = 2$ (solid), $\alpha = 1.5$ (dashed), $\alpha = 1$ (dotted). The CMB case corresponds to the solid gray line centered at 0. The vertical dotted line represents the maximum multipole ($\ell = 9000$) used in the analysis.}\label{fig:int_fg_sed}
 \end{figure}

Fig.~\ref{fig:bl_nu} shows the variation of a Gaussian $\blnu$ within three example channels at $\ch =$ 93, 145 and 225 GHz. Depending on the value of $\alpha$ considered and with a bandpass width of $\sim 30$ GHz, the chromaticity of the beam (see Eq.~\ref{eq:nu_scaling}) becomes more and more evident as we go to smaller scales.
Fig.~\ref{fig:int_fg_sed} shows the distortion of the foreground $a_{\lm}$s induced by a chromatic Gaussian beam as parametrized in Eqs.~\ref{eq:gaussbl} and \ref{eq:nu_scaling} for the same three frequency channels. In each panel of Fig.~\ref{fig:int_fg_sed}, we consider three beam chromaticity scalings, $\alpha = 1/1.5/2$, and three different extragalactic foregrounds, i.e., unresolved radio point sources, Cosmic Infrared Background (CIB) and thermal Sunyaev-Zeldovich (tSZ) (the explicit expressions of the SEDs of these components are reported in Appendix A of Ref.~\cite{Giardiello:2024uzz}). The change due to beam chromaticity is similar across frequencies and the dependence on the value of $\alpha$ similar across foreground components, with an overall distortion at the power spectrum level that grows at high-$\ell$s and that can be as large as $\sim20\%$ at the boundaries of our analysis, i.e., $\ell=9000$.

\section{Implementation and Forecast of impact on cosmological analyses} \label{sec:implementation}
To push this formalism to an example of a full cosmological analysis accounting for beam chromaticity, we implement this effect in the public SO power spectrum multi-frequency likelihood \texttt{LAT\_MFLike}, and in its  \texttt{fgspectra} library calculating the foreground SEDs and power spectra. To do this, we simply expand the modules to allow the integration in frequency as in Eq.~\ref{eq:clFG}. This implementation is not specific to the SO software, the beam chromaticity formalism shown here can be included in this way in any other CMB power spectrum likelihood which models foregrounds together with the pure CMB component. For example, in Appendix~\ref{app:bplike}, we compare the  \texttt{LAT\_MFLike} beam chromaticity implementation with the one present in the ACT \texttt{bplike}\footnote{\url{https://github.com/ACTCollaboration/bplike/tree/master}} code, showing agreement at numerical precision.

To quantify and highlight the potential impact of unaccounted-for beam chromaticity, we run an end-to-end analysis propagating this effect to the estimation of cosmological and astrophysical parameters with the Monte Carlo Markov Chain (MCMC) sampler \href{https://github.com/CobayaSampler/cobaya}{\texttt{Cobaya}}, using the modified \texttt{LAT\_MFLike} likelihood and assuming that no other systematics effect is present. 

In particular, we start by simulating an SO-like sky at at 93, 145 and 225 GHz generating a smooth CMB and foreground dataset (in power spectrum space for both temperature and E-mode of polarization) already deconvolved by the CMB beam. We follow closely the methodology and the analysis settings used in Ref.~\cite{Giardiello:2024uzz}. For the CMB theory we choose to simulate and explore a $\Lambda$CDM+$N_\mathrm{eff}$ model -- i.e., a single-parameter extension of the standard $\Lambda$CDM described by seven parameters: the baryon and dark matter densities, $\Omega_b h^2$ and  $\Omega_c h^2$, the amplitude and the spectral index of primordial scalar density perturbations, $A_s$ and $n_s$, both defined at a pivot scale $k_0=0.05$\,Mpc$^{-1}$, the Hubble constant, $H_0$ in km/s/Mpc, the optical depth to reionization, $\tau$, and the effective number of relativistic species, $N_\mathrm{eff}$. As well as allowing us to study specific relativistic particles such as neutrinos, $N_\mathrm{eff}$ is an ideal parameter to investigate the damping tail of the CMB where beam effects become important. 
The foreground power includes emission from: Galactic thermal dust, thermal and kinetic  Sunyaev-Zel’dovich effects (tSZ and kSZ), CIB clustered and Poisson terms, unresolved radio sources, and tSZ-CIB cross-correlation in temperature; Galactic dust and radio sources in polarization. The $\ell$ range considered is 30-9000, the passbands are top-hat in surface brightness units and with a $\sim 30$GHz width. The chosen values of all the cosmological and foreground parameters used in the simulation are the same as the benchmark smooth simulation of Ref.~~\cite{Giardiello:2024uzz}. For noise performance and simulated sensitivity of our dataset we use the SO public noise curves\footnote{The SO noise curves for temperature and polarization performance of the LAT are available at \href{https://github.com/simonsobs/so_noise_models/blob/master/so_models_v3/SO_Noise_Calculator_Public_v3_0_4.py}{\texttt{so\_models\_v3/SO\_Noise\_Calculator\_Public\_v3\_0\_4.py}}}. 
The total foreground power spectra are integrated with a chromatic beam as in Eq.~\ref{eq:clFG}, and as mentioned previously, beams are assumed to be Gaussian with different frequency scalings ($\alpha = 1/1.5/2$). We then estimate cosmological and foreground parameters assuming no chromatic effect, integrating the foreground emission as in Eq.~\ref{eq:clFGnochormaticity}.

Finally, we produce a benchmark case repeating this analysis using a smooth spectrum without beam chromaticity ($\alpha = 0$). We compare the results with chromatic beams to this simplified case to derive the mismatch induced on parameter posteriors. 

\begin{figure}[t!]
    \centering
    \includegraphics[width=0.8\textwidth]{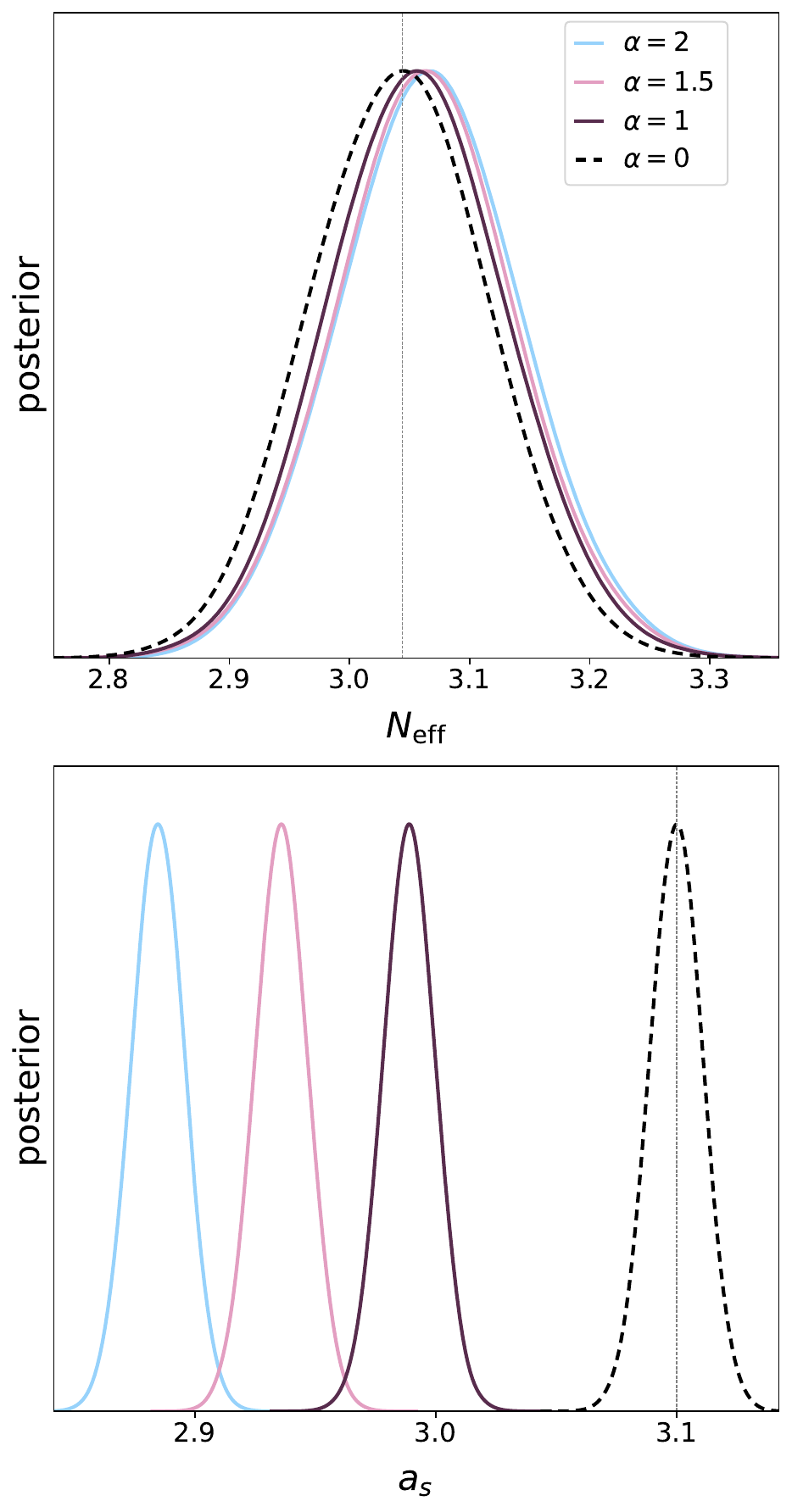}
    \caption{Shifts in the $N_\mathrm{eff}$ and $a_s$ (the amplitude of unresolved radio sources in temperature) posteriors due to the presence of Gaussian chromatic beams (with different frequency scaling $\alpha$) in the simulated spectra, not taken into account in the likelihood analysis. The case $\alpha=2$ represents the diffraction-limited beam for each spectral element, while $\alpha=0$ case represents the achromatic beam. The vertical dotted lines show the input values in the simulations which are recovered in the case of achromatic beams.}
    \label{fig:beam_cosmo_params}
\end{figure}

The key outcome of the runs done with these simulations is that neglecting the effect of beam chromaticity in the likelihood analysis primarily impacts the foreground parameters. In the case under study here, i.e., for an SO LAT-like experiment, all the $\alpha$ values considered cause shifts larger than 2$\sigma$ in the recovery of extragalactic foregrounds. We show an example of this with the posteriors of the radio source power in temperature, $a_s$, in the bottom panel of Fig.~\ref{fig:beam_cosmo_params}, for $\alpha = 1/1.5/2$ compared with the achromatic case. A summary of the impact on all the foreground parameters is reported in Appendix~\ref{app:tr_plots}. As expected, extragalactic components can be strongly biased, while diffuse Galactic foregrounds mainly dominant at large scales~\cite{Planck:2018yye,Planck:2018gnk} are less affected by main beam chromaticity. 

By construction, in our formalism the CMB power spectrum should remain unaffected by beam chromaticity, assuming that the ``CMB" beam is correctly estimated. Nevertheless, at small-scales the coupling between foregrounds and cosmological signals can mean that a misestimate of the foreground parameters will also induce biases in cosmological parameters. 
In our case, we observe biases up to 0.3$\sigma$ for $N_\mathrm{eff}$ and $H_0$ (and between 0.1$\sigma$-0.2$\sigma$ for the other cosmological parameters -- see Appendix~\ref{app:tr_plots}) when $\alpha = 2$. The top panel of Fig.~\ref{fig:beam_cosmo_params} shows that the bias in the recovery of $N_\mathrm{eff}$ increases with larger scalings. We do not find any degradation in the estimate of these parameters, i.e., the errors do not increase, we only see a rigid shift in the central values of the posteriors. 

We also tested the effect of unaccounted beam chromaticity in a cosmic-variance-limited survey with a resolution similar to that of the SO LAT. In this case, the distortion of the foreground spectra can induce biases on cosmological parameters up to several standard deviations. 

This analysis highlights the importance of the beam chromaticity for precise cosmology from future high-resolution measurements, especially to achieve accurate characterization of foregrounds, and to contain systematic biases on cosmological parameters below a small fraction of the error. Other systematic effects, e.g. errors in estimating the beam radial profiles or uncertainties in the passbands, can couple with beam chromaticity and potentially lead to even larger biases. These interactions, with a focus on SO, will be studied in a separate work.  

\section{Conclusions} \label{sec:conclusions}
In this paper, we have presented the formalism for modeling beam chromaticity and showed the impact of this higher-level correction of the beam in CMB power spectrum likelihood analysis. We showed how to implement this effect in the public SO software stack used for high-resolution multi-frequency CMB temperature and polarization analyses. We then quantified the impact of beam chromaticity on both cosmological and foreground parameters analyzing an SO-like case where the beam chromaticity effect is present in the simulated data but omitted in the likelihood analysis. We assumed no additional systematic effects to maintain simplicity and focus on testing the formalism introduced here. The chromatic beams, used in the simulations, were simulated as simple Gaussian beams with varying frequency scalings. We found biases at the level of several standard deviations on extragalactic foreground parameters, and up to $30\%$ of $\sigma$ on cosmological parameters, with the parameters most affected being those measured from the CMB damping tail. 
This demonstration shows the necessity of properly accounting for this effect in likelihood analyses of future high-resolution and high-sensitivity CMB data. A more thorough analysis for the SO LAT will follow, considering also the interplay between beam chromaticity and bandpass shifts and possibly the effect of marginalizing over the uncertainty on frequency scaling of the chromatic beam. An in-depth study of beam effects for the SO Small Aperture Telescopes is also underway~\cite{Dachlythra:inprep}. 

\begin{acknowledgments}
We acknowledge the use of \texttt{numpy} \citep{Harris:2020xlr}, \texttt{matplotlib} \citep{Hunter:2007} and \texttt{getdist} \citep{Lewis:2019xzd} software packages. We acknowledge the Hawk high-performance computing cluster at the Advanced Research Computing at Cardiff (ARCCA).
EC acknowledges support from the Horizon 2020 ERC Starting Grant (Grant agreement No 849169); SG and EC also acknowledge support from STFC and  UKRI (grant numbers ST/W002892/1 and ST/X006360/1). GG and LP~acknowledge the financial support from the COSMOS network (www.cosmosnet.it) through the ASI (Italian Space Agency) Grants 2016-24-H.0 and 2016-24-H.1-2018. The Flatiron Institute is supported by the Simons Foundation.
This is not an official Simons Observatory Collaboration paper.
\end{acknowledgments}

\bibliographystyle{apsrev4-1}
\bibliography{biblio}

\appendix

\section{Comparison between \texttt{LAT\_mflike} and \texttt{bplike} implementations of beam chromaticity}\label{app:bplike}

\begin{figure}[h!]
    \centering
    \includegraphics[width=1\textwidth]{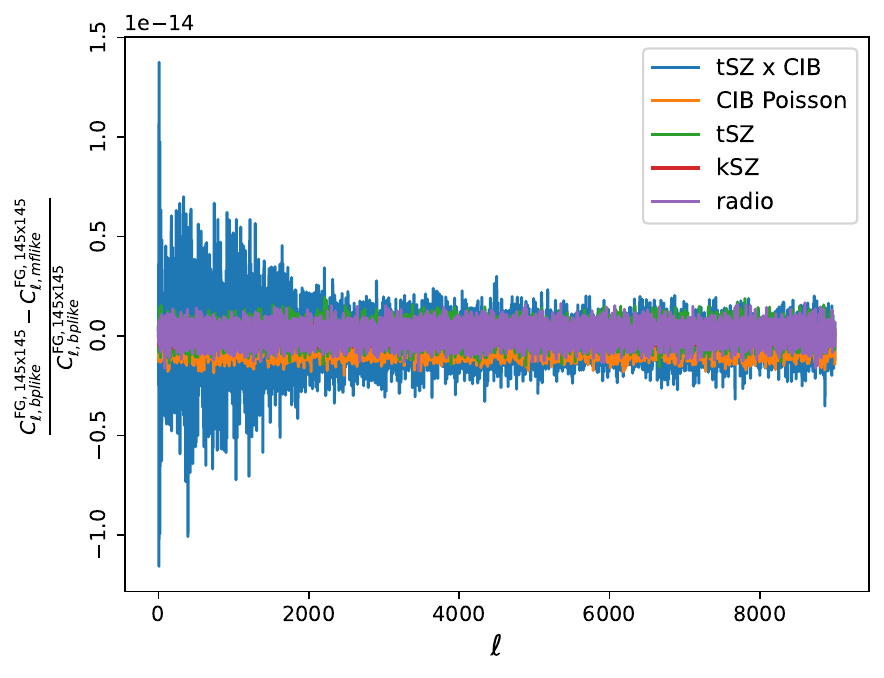}
    \caption{Relative difference between some extragalactic foreground spectra at 145~GHz, $C^{\nfg, 145, 145}_{\ell}$, computed with \texttt{LAT\_mflike} and \texttt{bplike} when using the same inputs. The spectra match at numerical precision level across the whole multipole range.}
    \label{fig:comparison}
\end{figure}

The implementation of beam chromaticity in \texttt{bplike} is similar to the one outlined in Section~\ref{sec:method}. The foreground power spectra are computed as 
\begin{align} \label{eq:C_tot_bfact}
&\hat{C}_{\ell}^{\nfg, \mathrm{c}_1 \mathrm{c}_2}=  \frac{\left[\int d\nu \, \tau^{\chp}_{\nu} \, \nfac_{\nu} b^{\chp}_{\ell, \nu}\,f^\mathrm{FG}_{\nu} \right]}{\left[  \int d\nu \, \tau^{\chp}_{\nu} \, \nfac_{\nu} \right] b^\mathrm{CMB, c_1}_{\ell}} \frac{\left[ \int d\nu \, \tau^{\chs}_{\nu} \, \nfac_{\nu} b^{\chs}_{\ell, \nu} \,f^\mathrm{FG}_{\nu}\right]}{\left[ \int d\nu \, \tau^{\chs}_{\nu} \, \nfac_{\nu}  \right] b^\mathrm{CMB, c_2}_{\ell}}C^{\nfg}_{\ell},
\end{align}
where the CMB beams are explicitly present. 
To compare and validate the two implementations, we compute the foreground power spectra of a single frequency channel, 145~GHz, with both codes and using the same input parameters, passbands and beams (the cosmological and foreground parameters and the passbands from our main analysis described in Section~\ref{sec:implementation}, and the case of beams scaled with $\alpha=2$). We do this for all the components modeled in the same way in the two codes: tSZ and kSZ, the Poisson component of the CIB, tSZxCIB, and radio sources (the Galactic dust SED is modeled as a modified black body in \texttt{LAT\_mflike} and as a power law in \texttt{bplike}, and there is a difference in the $\ell$ scaling of the CIB clustered component). All spectra match at numerical precision, as shown in Fig.~\ref{fig:comparison}. We expect similar agreement for other components once the SEDs and the template shapes in $\ell$ space have been aligned.

\FloatBarrier

\section{Cosmological and astrophysical parameters from simulations}\label{app:tr_plots}
\begin{figure*}
    \centering
    \includegraphics[width=0.8\textwidth]{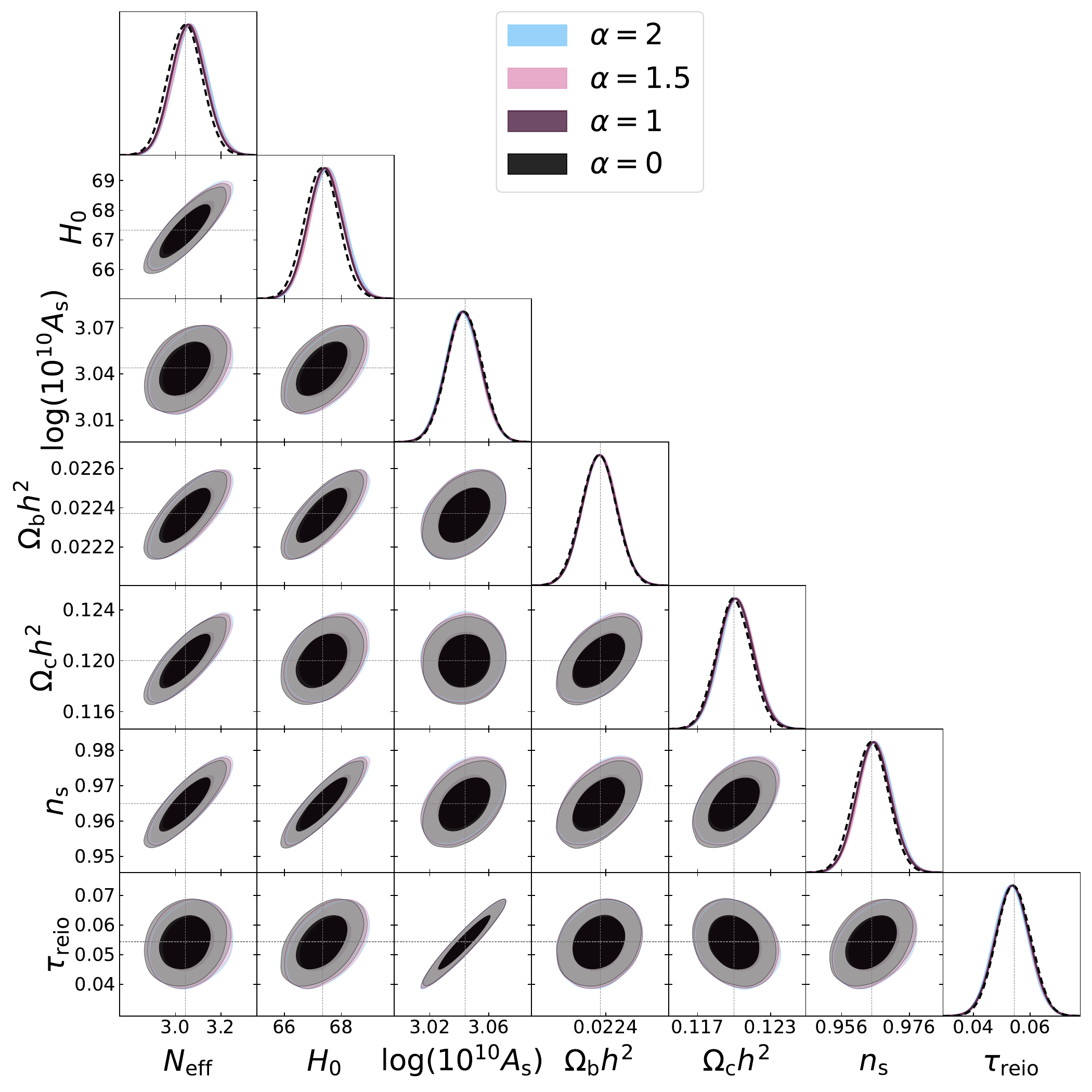}
    \caption{One- and two-dimentional distributions for the cosmological parameters recovered from simulations including Gaussian chromatic beams with different frequency scaling $\alpha$ and analysed without accounting for beam chromaticity in the likelihood. The case with $\alpha=2$ represents the diffraction-limited beam for each spectral element, while the $\alpha=0$ case represents the achromatic beam scenario. We see baises at the level of $0.2-0.3\sigma$ in parameters sensitive to the very small-scale region of the spectra where beams are very important. The vertical dotted lines show the input values in the simulations which are recovered in the case of achromatic beams.}
    \label{fig:tr_cosmo}
\end{figure*}

\begin{figure*}
\centering
    \includegraphics[width=1\textwidth]{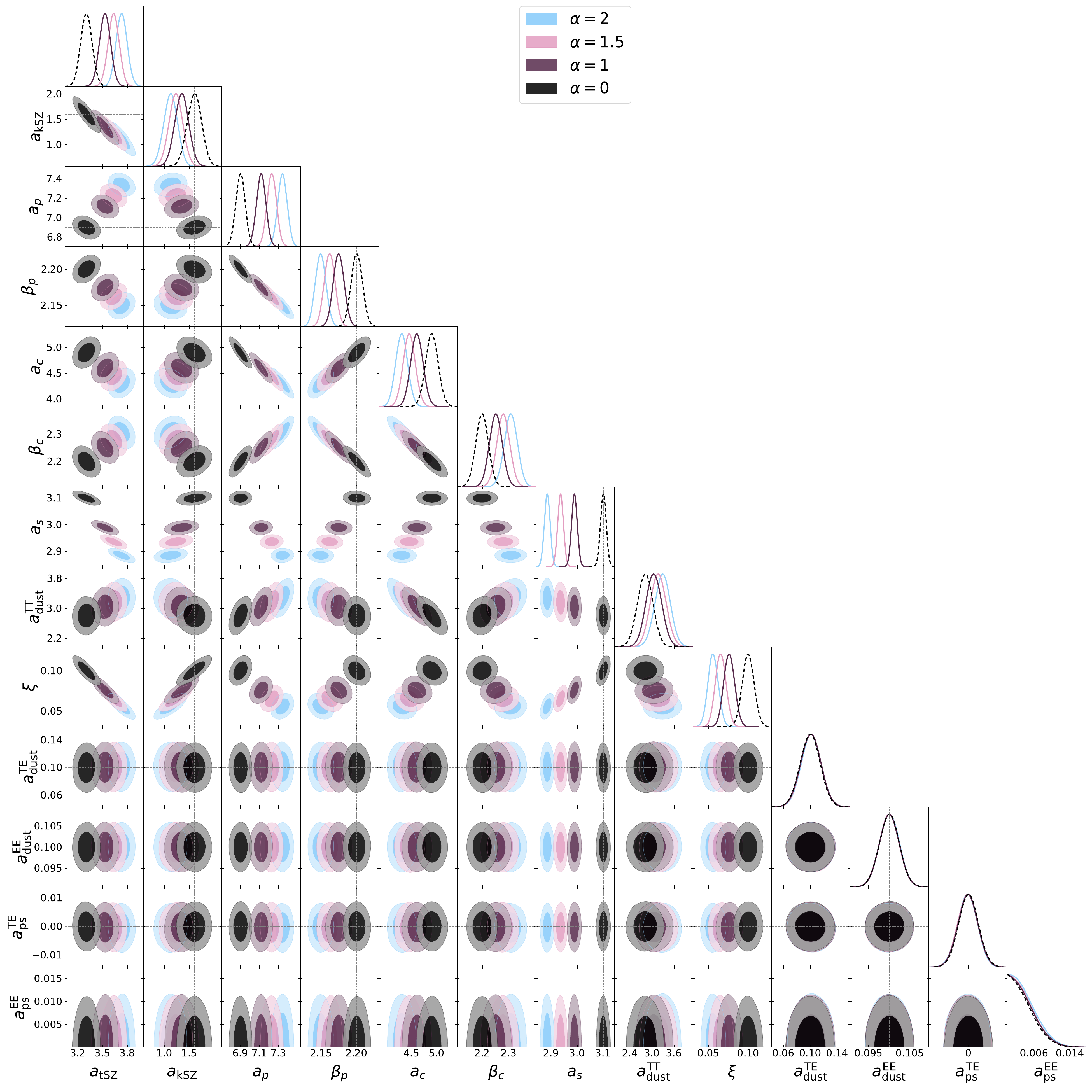}
    \caption{One- and two-dimentional distributions for the foreground parameters recovered from Gaussian chromatic beams with different frequency scaling $\alpha$ and analysed without accounting for beam chromaticity in the likelihood. The case $\alpha=2$ represents the diffraction-limited beam for each spectral element, while $\alpha=0$ case represents the achromatic beam. We see very large biases in the recovery of extragalactic foregrounds, and increasing with larger values of $\alpha=2$. The vertical dotted lines show the input values in the simulations which are recovered in the case of achromatic beams.}
    \label{fig:tr_fg}
\end{figure*}

We report below the full set of cosmological and foreground parameters estimated in the cases of different beam scalings. Figure~\ref{fig:tr_cosmo} and Fig.~\ref{fig:tr_fg} show the 1-dimentional posteriors and 2-dimentional contour levels for cosmology and astrophysics, respectively. They highlight that the impact on cosmology is at the level of fractions of their mesurement errors, while the shifts in astrophysics parameters can be very dramatic.

\end{document}